\documentclass[pra,aps,twocolumn,superscriptaddress,amsfonts,citeautoscript,nofootinbib,a4paper]{revtex4}

\usepackage{amsmath}  \usepackage{amssymb}  \usepackage{amsfonts}  \usepackage{bm}  \usepackage{bbm}   \usepackage{braket}  \usepackage{color}  \usepackage{comment}  \usepackage{dcolumn}  \usepackage{enumerate}  \usepackage{epsfig}  \usepackage{gensymb}  \usepackage{graphicx}  \usepackage{indentfirst}  \usepackage{lmodern}  \usepackage{mathrsfs}  \usepackage{mathtools}  \usepackage{psfrag}  \usepackage{pst-all}  \usepackage{soul}  \usepackage{units}  \usepackage{xcolor}
\usepackage[colorlinks,linkcolor=blue,citecolor=blue,urlcolor=blue,hyperindex,driverfallback=dvipdfm]{hyperref}  \usepackage[T1]{fontenc} 

\def\ii{{\rm i}}  \def\ee{{\rm e}}
\def\me{m_{\rm e}}  \def\kB{{k_{\rm B}}}
  
        \def\Eb{{\bf E}}        \def\Gb{{\bf G}}          \def\jb{{\bf j}}    \def\kb{{\bf k}}              \def\Rb{{\bf R}}  \def\rb{{\bf r}}      \def\vb{{\bf v}} 
\def\xx{\hat{\bf x}}                
\def\kpar{k_\parallel}  \def\kparb{{\bf k}_\parallel} 
\def\EF{{E_{\rm F}}}    \def\kF{{k_{\rm F}}} 
      
\def\ky{k_{y}} \def\kx{k_{x}}  \def\qx{q_{x}}    \def\qparb{{\bf q}_\parallel}  \def\EFz{{E_{\rm F}^0}}  \def\kFz{{k_{\rm F}^0}}

\begin{document} 

\title{Quantum-phase two-dimensional materials
}

\author{Valerio~Di~Giulio}
\affiliation{ICFO-Institut de Ciencies Fotoniques, The Barcelona Institute of Science and Technology, 08860 Castelldefels (Barcelona), Spain}

\author{P.~A.~D.~Gon\c{c}alves}
\affiliation{ICFO-Institut de Ciencies Fotoniques, The Barcelona Institute of Science and Technology, 08860 Castelldefels (Barcelona), Spain}

\author{F.~Javier~Garc\'{\i}a~de~Abajo}
\email{javier.garciadeabajo@nanophotonics.es}
\affiliation{ICFO-Institut de Ciencies Fotoniques, The Barcelona Institute of Science and Technology, 08860 Castelldefels (Barcelona), Spain}
\affiliation{ICREA-Instituci\'o Catalana de Recerca i Estudis Avan\c{c}ats, Passeig Llu\'{\i}s Companys 23, 08010 Barcelona, Spain}

\begin{abstract}
The modification of electronic band structures and the subsequent tuning of electrical, optical, and thermal material properties is a central theme in the engineering and fundamental understanding of solid-state systems. In this scenario, atomically thin materials offer an appealing platform because they are extremely susceptible to electric and magnetic gating, as well as to interlayer hybridization in stacked configurations, providing the means to customize and actively modulate their response functions. Here, we introduce a radically different approach to material engineering relying on the self-interaction that electrons in a two-dimensional material experience when an electrically neutral structure is placed in its vicinity. Employing rigorous theoretical methods, we show that electrons in a semiconductor atomic monolayer acquire a quantum phase resulting from the image potential induced by the presence of a neighboring periodic array of conducting ribbons, which in turn produces strong modifications in the optical, electrical, and thermal properties of the monolayer, specifically giving rise to the emergence of interband optical absorption, plasmon hybridization, and metal-insulator transitions. Beyond its fundamental interest, material engineering based on the quantum phase induced by noncontact neighboring structures represents a disruptive approach to tailor the properties of atomic layers for application in nanodevices.
\end{abstract}

\maketitle 
\date{\today} 

\section{Introduction}

At the heart of condensed-matter physics is the drive to manipulate materials in a purposeful fashion to improve or enable functionalities. To that end, the sustained advances in fabrication capabilities along with the continuous emergence of novel material platforms have together fuelled the field over the past half-century. Moreover, when suitably engineered, nanostructured materials can generally exhibit new properties beyond those found in their native, bulk form. A paradigmatic example that benefited from such approach was the development of semiconductor devices \cite{HV16,CG1985}, whose electronic properties were controlled by modifying the band structure through, for example, spatially patterning their compositional or doping characteristics \cite{ET1970,D1983,E1986,CG1985,SM1990,IP97}. 

With the advent of two-dimensional (2D) and atomically thin materials \cite{NJS05,GG13,NMC16,paper235}, those ideas were swiftly transferred to this arena as well, leading to the realization of 2D superlattices that incorporated not only vertical stacks, but also laterally assembled heterostructures \cite{YXC12,XTH18,ZZC18,FZW18,JGT19,LDF21}, as well as electrically modulated graphene \cite{CCO16,paper313}, artificial graphene \cite{WSD17,DLZ21}, and optical near-field dressing through periodic patterning of the supporting dielectric substrate \cite{XHZ21}. More recently, similar band structure engineering concepts have been explored to create moir\'{e} superlattices \cite{YXC12,DWM13,PG13,KCA17,YMJ19} and moir\'{e} excitons \cite{TMW19,JRY19,FBK21}, and also to investigate topological phenomena \cite{SSL15,CSF20,DHC21,paper298}. However, these approaches are generally invasive, as they require physical material nanostructuring or the injection of charge carriers. Now, the question arises, can a more gentle engineering of a 2D material be realized without structural modifications or exposure to external fields?

\begin{figure*}
\centering
\includegraphics[width=1.0\textwidth]{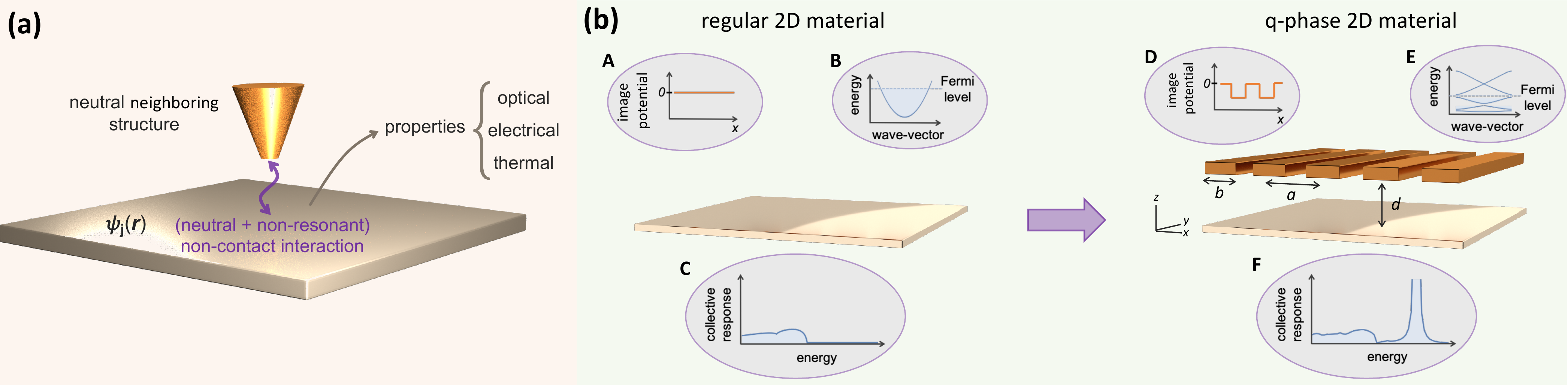}
\caption{{\bf Realization of a quantum-phase (Q-phase) material}. (a) Sketch of the noncontact interaction between a 2D material and a neighboring structure. The image potential experienced by conduction electrons in the material imprints a Q-phase on their wave functions $\psi_j(\rb)$ that in turn changes the optical, electrical, and thermal transport properties. (b) Possible realization of a Q-phase material. In the absence of an additional structure, there is no image interaction (A), so conduction electrons exhibit a characteristic parabolic dispersion (B), leading to a collective response function like that of a 2DEG (C). An image potential landscape (D) is produced by introducing a neighboring neutral structure (a periodic array of conducting ribbons of period $a$, width $b$, and separation $d$). The electron wave functions then acquire a Q-phase that reshapes the band structure, opening gaps (E) and enabling additional electronic transitions that translated into modifications of the material properties (F).}
\label{Fig1}
\end{figure*}

In this work, we introduce a disruptive approach for tailoring the electrical, optical, and thermal properties of 2D materials based on the manipulation of their electronic band structures by means of the gate-free, noncontact image-potential interaction experienced by the material's electrons in the presence of a neutral neighboring structure (see Fig.~\ref{Fig1}). Indeed, when a charged particle (e.g., an electron in the 2D material) is placed near an interface, an image potential is induced that affects the particle dynamics. For free electrons, the image interaction is tantamount to the position-dependent Aharonov--Bohm quantum phase (Q-phase) imprinted on their wave functions by the self-induced electromagnetic fields in the vicinity of the interface \cite{paper357}. Likewise, valence and conduction electrons in a material should acquire a Q-phase that depends on the geometrical and compositional details of the environment. This phase is expected to modify the electronic energy bands and, consequently, the dynamical response and transport properties of the hosting material as well. In reference to the origin of these modifications, hereinafter we refer to such 2D-material-based configurations as Q-phase materials.

Here, we introduce a specific realization of Q-phase materials consisting of a 2D semiconductor that is modified by the presence of a one-dimensional (1D) periodic array of ribbons. Specifically, we demonstrate that the periodic image potential produced by this pattern on the semiconductor electrons gives rise to substantial modifications in its band structure that translate into changes in the optical, electrical, and thermal properties, which are controlled by purely geometrical parameters (the separation and period of the array, see Fig.~\ref{Fig1}). We present rigorous theoretical calculations of the self-consistent electronic band structure, the ensuing optical response, and the electrical and thermal conductivities, all of which reveal dramatic modulations due to the aforementioned image interaction. More precisely, we predict the opening of electronic gaps, which enable interband optical transitions and hybrid plasmon-interband modes that are otherwise optically forbidden in the absence of the image interaction. In addition, we observe a metal-insulator transition as the pattern is brought closer to the semiconductor and the image interaction increases, essentially reflecting a reduction in carrier propagation induced by the periodic patterning. This also affects the thermal conductivity, which we predict to undergo a corresponding conductor-insulator transition. Importantly, the image interaction is long-range and nonresonant, so this method can be generally applied to any 2D material, adding a brand new tool to design nanodevices.


\section{Theoretical framework}

We consider an atomically thin semiconductor of area $L^2$, doped to a Fermi level $\EFz$ (corresponding to a carrier density $n_0$ in the conduction band), and lying in the $z=0$ plane at a distance $d>0$ from a 1D periodic array of conductive ribbons (period $a$, width $b$, see Fig.~\ref{Fig1}b). The latter are taken to be infinitely extended along the $y$ direction, periodic along $x$, and made of a material with a high DC conductivity and low infrared absorbance (e.g., indium tin oxide \cite{BF02}), so that it behaves as a perfect conductor with regards to the determination of the equilibrium configuration of the electronic structure in the semiconductor. We further consider the periodic array to be embedded in a medium of permittivity $\epsilon$ matching that of the ribbon material at the optical frequencies discussed below, such that the array appears to be invisible. The semiconductor electrons can however interact with the array through a image potential energy $-V_0$ \cite{JJ1988,FDB1972,E1981}, which for distances $d\gtrsim1$\,nm can be well approximated by the local electrostatic limit \cite{H1975,MRE1989,paper004} (i.e., $V_0=e^2/4\epsilon d$). Neglecting ribbon edge effects by assuming $d\ll a$, we describe the image interaction of each semiconductor electron through the periodic stepwise energy function $V^{\rm im}(x)=-V_0 p(x)$, where $p(x)=1$ for $x$ directly below a ribbon and $p(x)=0$ otherwise. In this work, we set $\epsilon=1$ for simplicity and remark that the image potential is the result of the Q-phase introduced in the electron wave functions due to their interaction with the surrounding patterned structure, as illustrated for free electrons moving near a material surface \cite{paper357}.

We describe the semiconductor electrons in the single-particle approximation \cite{HL1970}, and further consider them to be strongly confined to an atomic layer of small thickness compared with the separation $d$ from the conductive ribbons. This allows us to factorize the one-electron wave functions as $\psi_{\kb n}(\rb)=\psi_{\kparb n}(\Rb)\psi_\perp(z)$, with out-of-plane components yielding a probability density profile $|\psi_\perp(z)|^2\approx \delta(z)$. The remaining in-plane components depend on $\Rb=(x,y)$ and are determined by the self-consistent equation \cite{HL1970}
\begin{equation}
 \left[\mathcal{H}^0+V^{\rm im}+V^{\rm H}\right]\psi_{\kb_\parallel n}(\Rb)=\hbar\varepsilon_{\kb_\parallel n}\psi_{\kb_\parallel n}(\Rb), \label{SCE}
\end{equation}
where ${\mathcal{H}^0 = -\hbar^2 {\bm\nabla}_{\Rb}^2/2m^*}$ is the unperturbed electron Hamiltonian, $m^*$ denotes the effective mass, $V^{\rm im}$ is the aforementioned image potential, and $V^{\rm H}$ is the Hartree potential. Also, in virtue of Bloch's theorem, the electron eigenstates can be labeled with the in-plane wave vector $\kb_\parallel$ and the band index $n$, such that $\psi_{\kb_\parallel n}(\Rb)=\ee^{\ii \kb_\parallel \cdot \Rb}\, u_{\kb_\parallel n}(\Rb)/L$, where $u_{\kb_\parallel n}(\Rb)=u_{\kb_\parallel n}(\Rb+\ell a\xx)$ for any integer $\ell$. In particular, given the translational invariance of the system under consideration along the $y$ direction and the periodicity of the image potential along $x$, we have (see Appendix)
\begin{subequations}
\begin{align}
\psi_{\kparb n}(\Rb) &= \ee^{\ii \kb_\parallel \cdot \Rb}u_{\kx n}(x)/L , \label{eigenvec}\\
\varepsilon_{\kparb n} &= \varepsilon^x_{\kx n}+\hbar \ky^2/2m^*, \label{eigenval}
\end{align} 
\end{subequations}
where $u_{\kx n}$ and $\varepsilon^x_{\kx n}$ are directly obtained by projecting Eq.~(\ref{SCE}) on Fourier components (i.e., by expanding the potentials and the wave function in the form $f(x)=\sum_G f_G \ee^{\ii G x }$, where the sum extends over reciprocal lattice vectors $G$ that are multiples of $2\pi/a$, as detailed in the Appendix).

We solve Eq.\ (\ref{SCE}) iteratively by partially updating the Hartree potential ${V^{\rm H}(\Rb) = e^2 \int \text{d}^2 \Rb'\,\big[n(\Rb')-n_0\big]\big/|\Rb-\Rb'|}$ at each step, introducing the electron density $n(\Rb)=2\sum_{\kb_\parallel n} f_{\kparb n} |\psi_{\kparb n}(\Rb)|^2$ calculated from the (spin-degerate) electron wave functions and the Fermi-Dirac distribution $f_{\kparb n}$ (e.g., $f_{\kparb n}=\theta(\EF-\hbar \varepsilon_{\kparb n})$ at $T=0$). Importantly, we assume the semiconductor to remain electrically neutral in its environment, so that no charge imbalance is introduced by doping. Then, the charge neutrality condition $\int \text{d}^2\Rb ~ \big[n(\Rb)-n_0\big]=0$ needs to hold at every iteration step and is used to determine the Fermi energy $\EF$, which generally deviates from the value $\EFz$ in the homogeneous semiconductor (i.e., in the absence of image interaction). Incidentally, we note that the parabolic band assumed to describe the unperturbed conduction band of the doped semiconductor should be an excellent approximation because the patterning period $a$ exceeds by several orders of magnitude the interatomic distance, thus leading to a comparatively small 1D first Brillouin zone (1BZ) of extension $\sim 1/a$. 


\begin{figure*}
\centering
\includegraphics[width=1.0\textwidth]{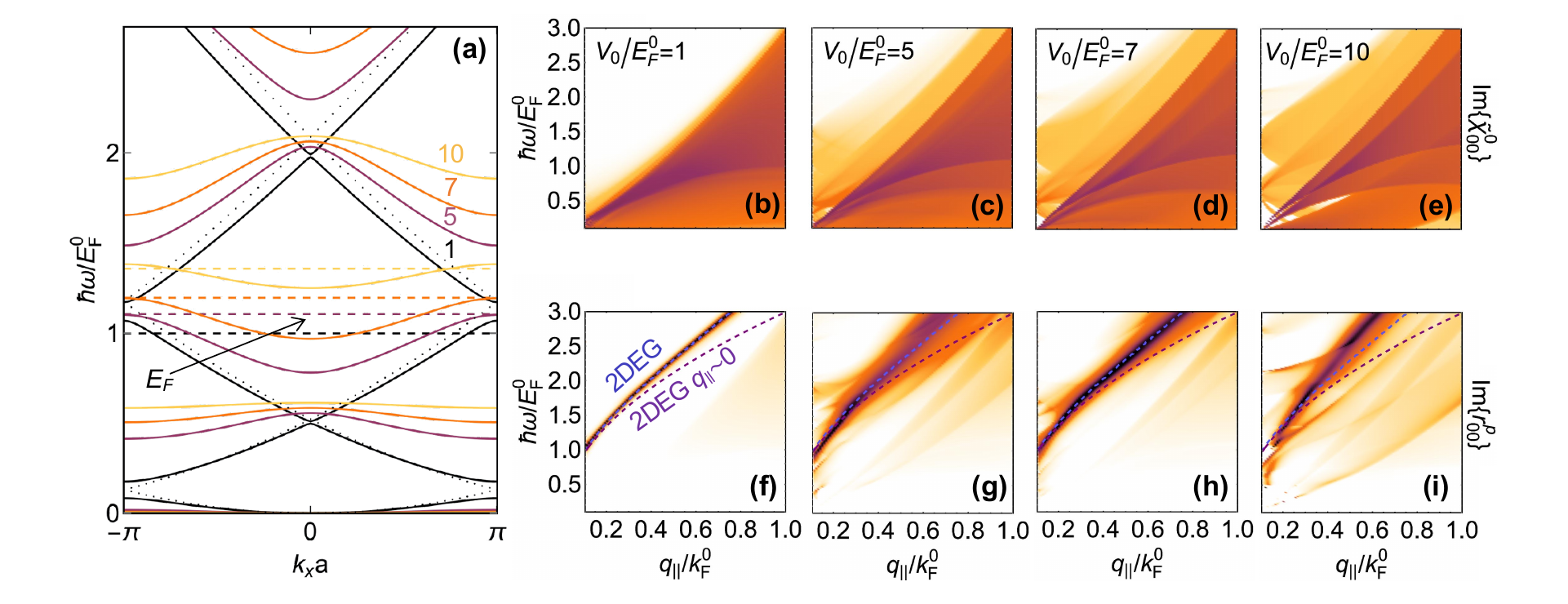}
\caption{\textbf{Modulation of the electronic band structure and optical response in a Q-phase material.} (a) Electronic bands (solid curves) in the periodic patterning direction for different strengths of the normalized image potential $V_0/\EFz$ (see color-matched labels), along with the corresponding normalized Fermi energies (dashed lines). Dotted curves stand for the 2DEG limit. (b-i) Single-particle excitations described by the 2D noninteracting susceptibility $\tilde{\chi}^0_{00}$ (b-e)  and collective response resonances revealed by the loss function ${\rm Im}\{r^{\rm p}_{00}\}$  (f-i) as a function of transferred energy ($\hbar\omega$) and in-plane wave vector ($q_\parallel=|(q_x,q_y)|$), normalized to the Fermi energy $\EFz$ and wave vector $\kFz$, respectively, for fixed $q_x/\kFz=0.1$. The plasmon dispersion relation in a 2DEG is shown for comparison in (f-i) (blue-dashed curves), along with its $q_\parallel/\kFz\ll 1$ limit (purple-dashed curves). All calculations are performed for $\EFz=0.29$\,eV, $m^*/\me=0.1$, $a=10$\,nm, and $b=5$\,nm.}
\label{Fig2}
\end{figure*}

\section{Modulation of the electronic band structure}

Once the conducting ribbons are brought close to the semiconductor, electrons in the latter are no longer free to move along the patterning direction $x$ because they are modulated by the self-induced potential resulting from the image interaction. In particular, the electronic band structure is modified by the emergence of band gaps at the centre and edges of the 1BZ (see Fig.~\ref{Fig2}a), which also imply changes in the electron velocity component $\partial_{k_x}\varepsilon^x_{k_x n}\equiv v_{k_xn}$ and the effective mass. We note that the normalized band energies $\hbar\varepsilon^x_{k_x n}/\EFz$ and the dimensionless eigenvectors $a\psi_{\kparb n}(\Rb/a)$ depend only on four independent parameters: $(i)$ the geometrical ratio $b/a$ (see Fig.~\ref{Fig1}b), which regulates the tunneling rate across barriers introduced by the image potential; $(ii)$ the size of the 1BZ $k_{\rm BZ}=\pi/a$ relative to the unperturbed Fermi wave vector (i.e., $k_{\rm BZ}/\kFz$); $(iii)$ the strength of the image energy relative to the unperturbed Fermi energy, $V_0/\EFz$; and $(iv)$ the normalized electron-electron Coulomb interaction energy across a unit cell $V_{\rm C}/\EFz$, where $V_{\rm C}=e^2/a$. As illustrated in the dispersion diagrams shown in Fig.~\ref{Fig2}a for different values of $V_0/\EFz$, when the ribbons are moved far apart, and thus $V_0/\EFz$ approaches 0, we rapidly recover a folded 2DEG (2D electron gas, dotted curves) at the center of the 1BZ, while small gaps of decreasing size remain visible at the zone edge. Reassuringly, we find that for vanishing $V_{\rm C}/\EFz$ the Hartree potential contributes negligibly to the energy of the system and the eigenvalues $\varepsilon^x_{\kx n}$ agree well with those obtained in the Kr\"{o}nig--Penney model \cite{KP1931}.

Here, we are interested in a regime where the image potential strongly affects the transport properties of the material. Such a regime is reached when the total image energy $E^{\rm im}$ dominates over the kinetic energy $E^{\rm kin}$, as otherwise the latter would push the system towards a ballistic behavior. The ratio between these two energies scales as $E^{\rm im}/E^{\rm kin}\sim V_0/\EFz$, assuming that the condition $(V_{\rm C}/\EFz)/(k_{\rm BZ}/\kFz)^2=(2/\pi^2)\,e^2m^*a/\hbar^2\gg1$ is satisfied (e.g., for $a\gg2.6$\,nm if $m^*=0.1\,\me$, see Appendix). This behavior is observed in the band structure calculations presented in Fig.~\ref{Fig1}a, where the influence of the image potential is visible through a monotonic increase in the band gaps with increasing strength of the image interaction $V_0$ (i.e., when bringing the ribbon pattern closer to the semiconductor). The reshaping of the semiconductor energy bands also produces a wealth of new dynamical and static properties which we investigate below.

 
\section{Optical response of Q-phase materials}

The modifications produced in the electronic band structure by the image interaction translate into substantial changes in the optical response. Given the symmetry of the system, we can work in the in-plane reciprocal space and separately deal with each 2D wave vector $\qparb$ (with $q_x$ within the 1BZ).  In addition, our structures involve small distances and periods compared with the light wavelength, so that the optical response can be safely described in the electrostatic limit. Consequently, we consider an external electric potential $\phi^{\rm ext}(\qparb,z,\omega)=\sum_G \phi_G^{\rm ext}(\qparb,z,\omega)\ee^{\ii(\qparb+G\xx)\cdot\Rb}$ acting on the Q-phase material at optical frequency $\omega$ and expanded in Fourier components labeled by 1D reciprocal lattice vectors $G$. Introducing a matrix notation, we can express the resulting induced potential in terms of the Coulomb interaction $\upsilon$ and the noninteracting 2D susceptibility $\tilde{\chi}^0$ as $\phi^{\rm ind}(\qparb,z,\omega)=\upsilon_G(\qparb,z)\cdot\tilde{\chi}^0(\qparb,\omega)\cdot[\mathcal{I}-\upsilon(\qparb,0)\cdot\tilde{\chi}^{0}(\qparb,\omega)]^{-1}\cdot\phi^{\rm ext}(\qparb,0,\omega)$, where dots indicate matrix multiplication, $\phi^{\rm ext}$ and $\phi^{\rm ind}$ are vectors of components $\phi_G^{\rm ext}$ and $\phi_G^{\rm ind}$, respectively, and we use the matrices $\mathcal{I}_{GG'}=\delta_{GG'}$, $\upsilon_{GG'}=\delta_{GG'}\upsilon_G$, and $\tilde{\chi}_{GG'}^{0}$. More precisely, the diagonal Coulomb matrix elements take the form $\upsilon_G(\qparb,z)=2\pi\,\ee^{-|\qparb+G\xx||z|}\big/|\qparb+G\xx|$, while we adopt the random-phase approximation (RPA) to evaluate the susceptibility from the one-electron eigenstates as \cite{HL1970,H1975} (see Appendix)
\begin{align}
&\tilde{\chi}_{GG'}^0(\qparb,\omega) \nonumber\\
&=\frac{e^2}{2\pi^2\hbar }\sum_{nn'}\int_{-\pi/a}^{\pi/a} d\kx \,M^{nn'}_G(\kx,\qx)[M_{G'}^{nn'}(\kx,\qx)]^* \nonumber\\
\times &\int_{-\infty}^\infty d\ky \frac{f_{\kparb-\qparb,n'}-f_{\kparb n}}{\omega -\varepsilon_{\kparb n}+\varepsilon_{\kparb-\qparb n'}+\ii0^+}, \label{chi0main}
\end{align}
where we have introduced the matrix elements $M^{nn'}_G(\kx,\qx)=(1/a)\int_0^a dx\,\ee^{-\ii G x}\,u^*_{\kx-\qx,n'}(x)\,u_{\kx n}(x)$.

The energy differences in the denominator of Eq.~(\ref{chi0main}) correspond to one-electron excitations, which show up as intraband ($n=n'$) and interband ($n\neq n'$) transitions in the plots of $\tilde{\chi}_{GG'}^0(\qparb,\omega)$ presented in Fig.~\ref{Fig2}b-e as a function of photon energy $\hbar \omega$ and parallel wave vector $q_\parallel$ for the $G=G'=0$ component and different strengths of the image interaction $V_0$. At low $V_0$, the dispersion diagram is dominated by the intraband excitation region that characterizes the conduction band of a doped semiconductor (i.e., the homogeneous 2DEG limit). As we increase $V_0$ (see Fig.~\ref{Fig2}b-e), the gap openings discussed above (Fig.~\ref{Fig2}a) enables interband excitations, and in particular, vertical transitions become available due to the proximity of the ribbon array.

The RPA accounts for the self-consistent interaction with the induced potential, thereby resulting in collective electron excitations. To explore these so-called plasmons, we calculate the Fresnel reflection coefficient for p-polarized waves $r^{\rm p}$, which relates the induced and external potentials via \cite{paper349} $\phi_G^{\rm ind}(\qparb,0,\omega)=-\sum_{G'}r_{GG'}^{\rm p}(\qparb,\omega)\,\phi_{G'}^{\rm ext}(\qparb,0,\omega)$. Using the formalism outlined above, the Fourier components of $r^{\rm p}$ are given by  
\begin{align}
r^{\rm p}_{GG'}(\qparb,\omega)=\frac{1}{1- \big[\upsilon(\qparb,0)\cdot\tilde{\chi}^0(\qparb,\omega)\big]^{-1}}\Bigg|_{GG'}. \nonumber
\end{align}
In Fig.~\ref{Fig2}f-i, we plot the loss function ${\rm Im}\{r^{\rm p}_{GG'}\}$ obtained from this equation as a function of $q_\parallel$ and $\omega$. We note that even though we concentrate on the specular-reflection coefficient corresponding to $G=G'=0$, the condition $k_{\rm BZ}\ll\kFz$ requires the evaluation of $G$ components up to $G\gg\kFz$ in the $\upsilon$ and $\tilde{\chi}^0$ matrices to correctly account for transitions happening close to the Fermi surface.

Collective plasmon excitations are identified as intense features in Fig.~\ref{Fig2}f-i, which should be measurable through electron energy-loss spectroscopy, as we discuss in the Appendix (see Fig.~\ref{FigS1}). For relatively  small $V_0/\EFz$ (Fig.~\ref{Fig2}f), the dispersion diagram is dominated by a single, continuous plasmon band, in excellent agreement with the plasmon dispersion of the textbook uniform 2DEG (superimposed). This agreement reflects the fact that the plasmon behavior is mainly controlled by the average electron density $n_0$, provided the external perturbation produced in the band structure by the periodic image potential is still weak (see the relatively small gap openings in Fig.~\ref{Fig2}a for $V_0=\EFz$). Incidentally, at low $V_0$ the plasmon is qualitatively well described by the dispersion relation $\omega_{\rm p}(q_\parallel)\sim e \sqrt{n q_\parallel/m^*}$ obtained in the $q_{\parallel}/\kFz\ll 1$ limit. In contrast, as the ribbon structure is brought closer to the semiconductor, so that the image potential energy increases and eventually dominates in the system, a zoo of excitations emerge in the dispersion diagram: besides the 2DEG plasmon, features associated with interband transitions and their hybridization with plasmons are revealed. In addition, all of these features are dressed by electron-electron interactions, leading to a blue shift of the plasmon relative to the 2DEG limit, as well as spectral shifts of the interband transitions relative to the undressed excitations depicted in Fig.~\ref{Fig2}b-e.

\begin{figure}
\centering \includegraphics[width=0.4\textwidth]{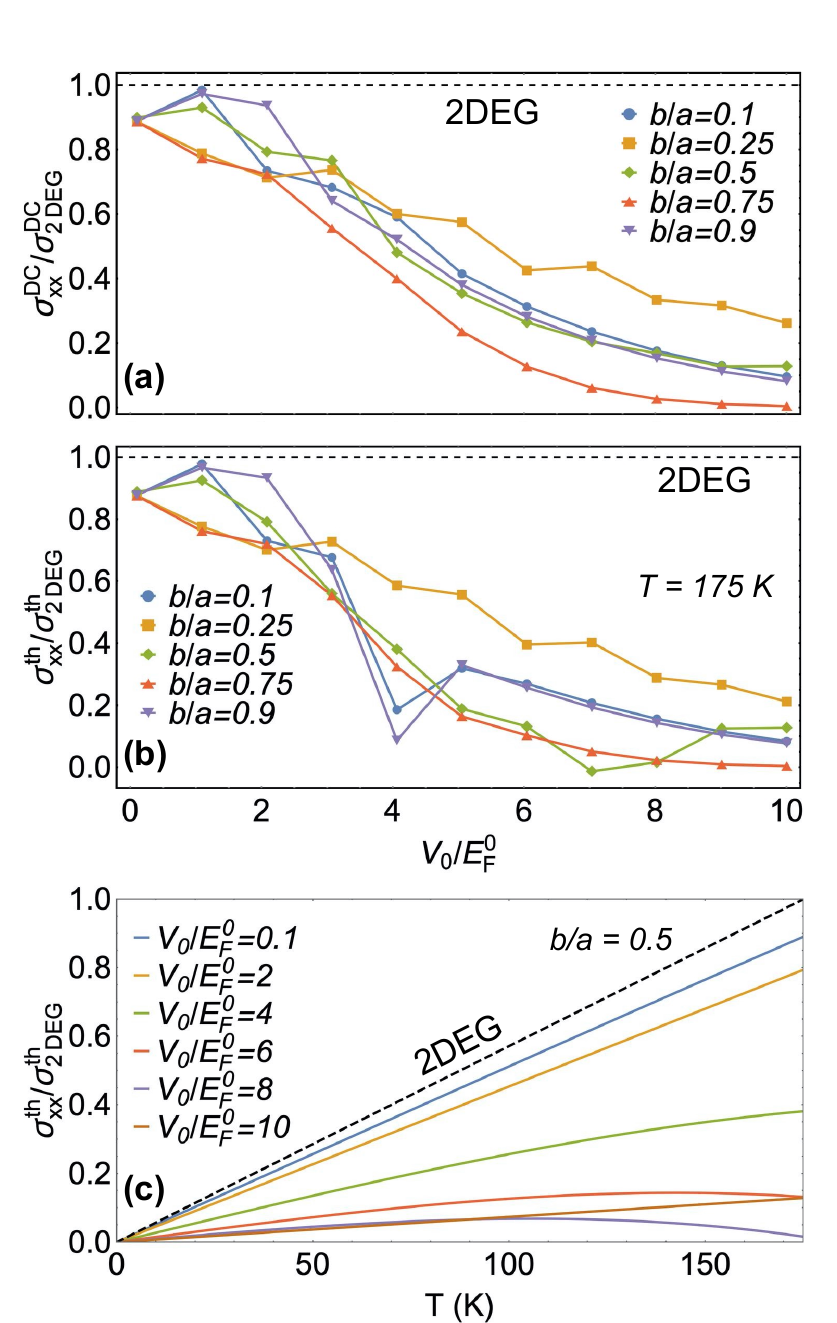}
 \caption{\textbf{Q-phase modulation of the DC electrical and thermal conductivities.} (a) Component of the 2D electrical conductivity tensor along the periodicity direction as a function of the normalized image potential strength $V_0/\EFz$ for several values of the $b/a$ ratio (see legend). (b) Thermal conductivity at $T=175$\,K normalized to that of a 2DEG under the same conditions as in (a). (c) Temperature dependence of normalized thermal conductivity for $b/a=0.5$ and different values of $V_0/\EFz$ (c). We use the same material parameters as in Fig.~\ref{Fig2}.}
\label{Fig3}
\end{figure}

\section{Metal-insulator transition}

Beyond the optical response, we expect the image interaction to also modify the static properties of Q-phase materials, including the DC electrical conductivity $\sigma^{\rm DC}$. We compute this quantity in the relaxation-time approximation \cite{AM1976}, introducing a phenomenological inelastic scattering time $\tau$ (see Appendix), so the conductivity is uniquely determined by the band energies $\hbar\varepsilon_{\kparb n}$ and the Fermi energy $\EF$ (Fig.~\ref{Fig2}a). Because of the symmetry of the system, the $2\times2$ conductivity tensor should only contain diagonal $xx$ and $yy$ components (i.e., along directions parallel and perpendicular to the periodic modulation). In addition, the $yy$ component remains unchanged with respect to the unperturbed semiconductor (i.e., $\sigma_{yy}^{\rm DC}=e^2\EFz\tau/\pi\hbar^2$) because the $x-$averaged electron density is conserved. The conductivity along $x$ can then be directly computed from the energy distribution in Eq.\ (\ref{eigenval}). In particular, for sufficiently large electronic gaps compared with $\kB T$, the DC electrical conductivity can be well described by its zero-temperature version  
\begin{align}
\sigma^{\rm DC}_{xx}&=
(e^2m^* \tau/\pi^2\hbar^2)\sum_n \int_{-\pi/a}^{\pi/a} d\kx \, ( \partial_{\kx}\varepsilon^x_{\kx n} /\Delta\kx),\label{sigdc}
\end{align}
where $\Delta \kx= \hbar^{-1}\big[2m^*(\EF-\hbar \varepsilon^x_{\kx n})\big]^{1/2}$. We use this expression in combination with the bands plotted in Fig.~\ref{Fig2}a to obtain the results presented in Fig.~\ref{Fig3}. Remarkably, the thermal conductivity exhibits a steady decrease with increasing image interaction $V_0$, starting from the 2DEG value in the unperturbed semiconductor at $V_0=0$ and evolving towards a substantial suppression when $V_0$ is a multiple of $\EFz$. This behavior is a direct consequence of the reduction in the ability of charge carriers to move in the periodic potential landscape produced by the image interaction. We thus predict a metal-insulator transition as the latter is switched on by placing the conductive ribbons closer to the semiconductor (i.e., by increasing $V_0$). This transition occurs for all ribbon sizes under consideration, with an optimum behavior found for $b/a\sim0.75$.

\section{Inhibition of the thermal conductivity}

Similar to the electron conductivity, the electronic thermal conductivity $\sigma^{\rm th}$ undergoes strong modifications due to the image interaction, since it is equally mediated by carrier propagation. We adopt again the relaxation-time approximation \cite{AM1976} and compute $\sigma^{\rm th}$ from the electronic band structure (see Appendix). Following similar arguments as above, we find the conductivity tensor to be diagonal, with its $yy$ component $\sigma_{yy}^{\rm th}=(\pi\kB^2T\tau/9\hbar^2)\big[3\EFz-(\pi\kB T)^2\big]$ being unaffected by the image interaction. Calculating the remaining $xx$ component up to third order in $\kB T$, we find
\begin{align}
\sigma^{\rm th}_{xx} &=\frac{\pi^2k_{\rm B}^2T\,\sigma_{xx}^{\rm DC}}{3e^2}\,\left[1 - \left(\frac{\pi }{3}\,k_{\rm B} T\,\frac{\sigma_{xx}^\prime}{\sigma_{xx}^{\rm DC}}\right)^2\right]\label{sigth},
\end{align}
where $\sigma_{xx}^{\rm DC}$ is the DC electrical conductivity in Eq.\ (\ref{sigdc}) and $\sigma'_{xx,}=(e^2m^* \tau /\pi^2\hbar^3)\sum_n \int_{-\pi/a}^{\pi/a} d\kx \, (\partial_{\kx}^2 \varepsilon^x_{\kx n}/\Delta\kx)$. The $V_0$-dependent thermal conductivity plotted in Fig.~\ref{Fig3}b reveals a transition from a thermal conductor to a thermal insulator analogous to the electrical behavior, which we also attribute to the reduction in carrier propagation produced by the periodic image potential. In addition, we observe a clear departure from the Wiedemann--Franz law \cite{FW1853} when comparing Figs.~\ref{Fig3}a and \ref{Fig3}b, as revealed by the oscillations displayed by $\sigma^{\rm th}_{xx}$ as a function of $V_0$, which are absent in $\sigma^{\rm DC}_{xx}$. Incidentally, the thermal conductor-insulator transition is now faster at lower values of $b/a$. This transition is further explored in Fig.~\ref{Fig3}c as a function of temperature for $b/a=0.5$, where a sustained reduction in the thermal conductivity with increasing $V_0$ is found to take place over a wide range of temperatures, leading to a large departure from the characteristic linear regime of the 2DEG.

\section{Conclusions}

In conclusion, we have demonstrated, based on rigorous theory, that the optical, electrical, and thermal properties of a 2D material can be substantially modified by introducing a neutral, noncontact structure in its vicinity. This constitutes a genuinely radical departure from currently available methods to engineer material properties, as instead we capitalize on the image interaction, which translates into a quantum phase imprinted on the valence electrons of the material. We remark the nonresonant nature of such interaction, which should therefore be generally applicable to modulate the properties of different types of materials, provided their thickness is small enough as to be strongly influenced by the image interaction with the added structure. As a possible realization, we have studied the modification in the properties of a 2D semiconductor when an array of conductive ribbons is brought in close proximity. Specifically, energy gaps are induced in the electronic band structure, interband electronic transitions are enabled, a rich landscape of additional plasmon bands emerges, and the electrical/thermal conductivity displays a metal/conductor-insulator transition as the semiconductor-array distance is reduced and the image interaction is increased. Our results represent a first step towards the realization of gate-free material tunability, potentially granting access into a whole range of properties that could find application in the design of nanodevices.

\appendix 
\section*{Appendix}



\section{Self-consistent one-electron wave functions}\label{meth1}

We note that Eq.~(\ref{SCE}) can be solved by factorizing the one-electron wave functions as $\psi_{\kparb n}(\Rb)=\psi_{\kx n}(x) \varphi_{\ky}(y)$, leading to
\begin{subequations}
\begin{align}
&\Big(\frac{\hbar^2}{2m^*}\partial_y^2+\hbar\varepsilon^y_{\ky n}\Big)\,\phi_{\ky}(y)=0, \label{1schy} \\
&\Big[ \frac{\hbar^2}{2m^*}(\kx-\ii \partial_x)^2 +V_{\rm im}(x)+V_{\rm H}(x)-\hbar \varepsilon^y_{\kx n}\Big]\,u_{\kx n}(x)=0,\label{1dreal}
\end{align}
\end{subequations}
where we have used Bloch's theorem to express $\psi_{\kx n}(x)=\ee^{\ii \kx x} u_{\kx n}(x)/\sqrt{L}$ in terms of a periodic function $u_{\kx n}(x)$ with the same period $a$ as the ribbon array. From Eq.\ (\ref{1schy}), the component along the direction of translational invariance admits plane-wave solutions $\phi_{\ky}(y)=\ee^{\ii \ky y}/\sqrt{L}$ with parabolic dispersion. In the remaining in-plane direction, Eq.\ (\ref{1dreal}) transforms into an eigenvalue problem by expanding $u_{\kx n}(x)=\sum_G \ee^{\ii G x} u_{\kx n,G}$ as a sum over reciprocal lattice vectors $G$ (multiples of $2\pi/a$). More precisely, Eq.\ (\ref{1dreal}) becomes
\begin{align}
\Big[\frac{\hbar^2}{2m^*}(\kx+G)^2-&\hbar \varepsilon^x_{\kx n} \Big]\,u_{\kx n,G} \nonumber\\
+\sum_{G'} &\Big(V_{G-G'}^{\rm im}+V^{\rm H}_{G-G'}\Big)\,u_{\kx n,G'}=0, \label{1df}
\end{align}  
where $V_G^{\rm im}=\ii\,\big(V_0/a G\big)\big[1-\ee^{-\ii G b}\big]$ and $V_G^{\rm H}=2\pi e^2(1-\delta_{G,0})n_G\big/|G|$ are the Fourier components of the image and Hartree potentials, respectively (see Appendix). Finally, we solve Eq.~(\ref{1df}) iteratively by calculating the components of the electron density $n_G=(2/a)\int_0^a dx\,\ee^{-\ii Gx}\sum_{\kb_\parallel n} f_{\kparb n} |\psi_{\kparb n}(\Rb)|^2$ at every step and adjusting the Fermi energy $\EF$ to meet the condition that $n_{G=0}$ is equal to the unperturbed electron density $n_0$.

\section{Calculation of single-particle electron states \label{subsec1}}

We consider a semiconductor monolayer under the conditions described in the main text, characterized by a parabolic conduction band of effective mass $m^*$, partially filled to a Fermi level $\EFz$. Conduction electrons can then be labeled by the in-plane wave vector $\kparb=(k_x,k_y)$, such that the electron energies are $\hbar\varepsilon_{\kparb}=\hbar^2\kpar^2/2 m^*$ relative to the bottom of the conduction band. We also incorporate a periodic image potential produced by interaction with a neighboring conductive ribbon array of parameters and material composition as described in the main text (see Fig.~\ref{Fig1}b). The image potential landscape is introduced through a term $V^{\rm im}(x)=-V_0 p(x)$ in the electron Hamiltonian, where $V_0>0$ measures the magnitude of the image interaction and $p(x)=\sum_{\ell=-\infty}^\infty \theta(\ell a+b-x) \theta(x-\ell a)$ follows the profile of the ribbon array (width $b$, period $a$) by means of step functions. Electron bands are then emerging, so we restrict $k_x$ to the 1BZ and introduce a band index $n$ to label electron states by $(\kparb,n)$ with $|k_x|<\pi/a$. We only consider in-plane electron motion (i.e., in the $\Rb=(x,y)$ plane), under the assumption that electron states are tightly confined along the out-of-plane direction, such that their wave functions can be factorized as $\psi_{\kparb n}(\Rb)\psi_\perp(z)$, where $\psi_\perp(z)$ is shared by all states. In addition, we approximate the out-of-plane probability density as $|\psi_\perp(z)|^2\approx \delta(z)$. The remaining in-plane components are governed by the Schr\"{o}dinger equation 
\begin{align}
\mathcal{H}(\Rb)\psi_{\kparb n}(\Rb) = \hbar \varepsilon_{\kparb n}\psi_{\kparb n}(\Rb), \label{schreal} 
\end{align}
with the Hamiltonian
\begin{align}
\mathcal{H}(\Rb)&=\mathcal{H}^0(\Rb) + V^{\rm im}(x) + V^{\rm H}(\Rb), \label{hami}
\end{align}
where $\mathcal{H}^0(\Rb)=-\hbar^2\nabla^2_{\Rb}/2m^*$ and we introduce electron Coulomb repulsion through the Hartree potential \cite{HL1970} $V^{\rm H}(\Rb)=e^2 \int d^2 \Rb' \big[n(\Rb')-n_0\big]/|\Rb-\Rb'|$, where $n(\Rb)=2\sum_{\kparb n} f_{\kparb n}|\psi_{\kparb n}(\Rb)|^2$ is the electron density, the factor of 2 accounts for spin degeneracy, and $f_{\kparb n}$ is the Fermi--Dirac distribution. Here, we calculate the electronic band structure at zero temperature, such that $f_{\kparb n}=\theta(\EF-\hbar\varepsilon_{\kparb n})$, where the Fermi energy $\EF$ is adjusted to make the average electron density equal to that of the unperturbed 2D semiconductor $n_0$. Then, $\EF$ depends on the applied image potential and generally differs from $\EFz$. We solve Eq.\ (\ref{schreal}) iteratively by calculating the Hartree potential at each step, fixing $\EF$ to preserve the average electron density, and mixing the new Hartree potential with the preceding one until convergence is achieved after a few iterations.

Because of the lattice periodicity of the Hamiltonian (i.e., $\mathcal{H}(\Rb)=\mathcal{H}(\Rb+\ell \xx)$ for any integer $\ell$), the eigenstates can be written as $\psi_{\kparb n}(\Rb)=\ee^{\ii \kparb \cdot \Rb}u_{\kparb n}(\Rb)/L$ (Bloch's theorem), where $L^2$ is the semiconductor area, and the functions $u_{\kparb n}(\Rb)$ also satisfy $u_{\kparb n}(\Rb+\ell \xx)=u_{\kparb n}(\Rb)$ for any integer $\ell$. In addition, since the image potential only depends on $x$, we can factorize the wave functions and separate the energies as
\begin{subequations}
\begin{align}
\psi_{\kparb n}(\Rb)&=\ee^{\ii \kparb \cdot \Rb}u_{\kx n}(x)/L,\label{eigenve}\\
\hbar\varepsilon_{\kparb n}&=\hbar \varepsilon^x_{\kx n}+\hbar^2\ky^2/2m^*, \label{eigenva}
\end{align}
\end{subequations}
where we have plane waves in the direction of translational invariance $y$. Along the direction of periodicity $x$, we find $\varepsilon^x_{\kx n}$ and $u_{\kx n}(x)$ by solving the 1D problem 
\begin{align}
&\left[ \hbar^2(\kx-\ii \partial_x)^2/2m^*+V^{\rm im}(x)+V^{\rm H}(x)\right]  u_{\kx n}(x)
\nonumber\\
&=\hbar \varepsilon_{\kx n} u_{\kx n}(x), \label{1dschreal}
\end{align}
where we introduced the 1D Hartree potential
\begin{align}
V^{\rm H}(x)=-2e^2\int_{-\infty}^\infty dx'\, \big[n(x')-n_0\big]\,\log|x-x'|, \label{1dhart}    
\end{align}
and $n(x)=(1/\pi^2)\sum_{n}\int_{-\pi/a}^{\pi/a}d\kx  \int_{0}^\infty d\ky\,  |u_{\kx n}(x)|^2\,\theta(\EF-\hbar\varepsilon^x_{\kx n}-\hbar^2\ky^2/2m^*)$ is the electron density profile. To obtain Eq. (\ref{1dhart}), we have employed the prescription $\sum_{\kparb} \rightarrow (L/2\pi)^2\int d^2\kparb $, used the integral and limit $\int_0^b dy/\sqrt{(x-x')^2+y^2}=\log[b+\sqrt{(x-x')^2+b^2}]-\log|x-x'|\xrightarrow[b\to\infty]{}\log(2b)-\log|x-x'|$, and applied the condition of charge neutrality to eliminate $x$-independent terms. Now, by moving to Fourier space, we transform Eq. (\ref{1dschreal}) into a linear system of equations,
\begin{align}
&\left[\frac{\hbar^2}{2m^*}(\kx+G)^2-\hbar \varepsilon^x_{\kx n} \right]u_{\kx n,G} \nonumber\\
&+\sum_{G'}\left(V_{G-G'}^{\rm im}+V^{\rm H}_{G-G'}\right)u_{\kx n,G'}=0, \label{eqrec}
\end{align}
where $G$ and $G'$ are reciprocal lattice vectors (i.e., multiples of $2\pi/a$). This equation involves the Fourier coefficients of different quantities, defined through the relations $f_G=(1/a)\int_0^a dx\,f(x)\ee^{-\ii Gx}$ and $f(x)=\sum_G f_G\ee^{\ii G x}$. In particular, from the stepwise profile of the image potential (see above), we find 
\begin{align}
V_G^{\rm im}&=V_0\times\begin{cases} ~(\ii/aG)\,\left(1-\ee^{-\ii G b}\right), &{\rm for}~G\neq 0,\\
-b/a, &{\rm for}~G= 0.\end{cases} \nonumber
\end{align}
Likewise, from Eq. (\ref{1dhart}), the coefficients of the Hartree potential reduce to
\begin{align}
V_G^{\rm H}&=2\pi e^2\times\begin{cases} n_G/|G|,~&{\rm for}~G\neq 0,\\
0, &{\rm for}~G= 0, \end{cases}\nonumber
\end{align}
where we have used the integral $\int_{-\infty}^\infty dx \,\ee^{ \ii G x} \log|x|=-\pi/|G|$ (see Eq. (4.441-2) in Ref. \cite{GR1980}).

\section{Quantification of the image energy}

We intend to quantify the influence of the image interaction on the electronic behavior in the semiconductor considering the different parameters that define the system. We start from the density $n=k_{\rm F}^2/2\pi$ and the kinetic energy $E^{\rm kin}=L^2\pi\hbar^2n^2/2m^*$ of a 2DEG. Following a density functional theory approach in the local-density approximation, we write the total energy  as a functional of the electronic density $n(\Rb)$:
\begin{align}
E[n]=&\frac{\pi \hbar^2}{2m^*}\int d^2 \Rb\, n^2(\Rb)+\int d^2\Rb\,V^{\rm im}(\Rb)n(\Rb) \nonumber\\
    &+\frac{e^2}{2}\int d^2\Rb d^2 \Rb'\, \frac{ \Delta n(\Rb)\Delta n(\Rb')}{|\Rb-\Rb'|}, \label{totenergy}
\end{align}
where $\Delta n(\Rb)=n(\Rb)-n_0$. The ground-state density in the many-electron system is then obtained by minimizing Eq.\,(\ref{totenergy}), subject the constraint $\int d^2 \Rb\, \Delta n(\Rb)=0$. By introducing a Lagrange multiplier $\lambda$ and imposing the vanishing of the functional derivative with respect to $n$ (i.e., $\delta_n E[n]=\lambda$), we find
\begin{align}
\frac{\pi\hbar^2}{m^*} [\Delta n(\Rb)+n_0]&+V^{\rm im}(x) +\frac{e^2}{2}\int d^2\Rb' \frac{\Delta n(\Rb')}{|\Rb-\Rb'|}=\lambda,\nonumber
\end{align}
which, transforming all quantities to reciprocal space as in Sec. (\ref{subsec1}), can be written as 
\begin{align}
\frac{\pi \hbar^2}{m^*}\left[\Delta n(x)+n_0\right]&+V^{\rm im}(x) \label{lagmult}\\
&+e^2 \pi \sum_{G\neq 0} \frac{\Delta n_G}{|G|}\ee^{\ii G x}=\lambda. \nonumber
\end{align}
Now, as a crude approximation, we assume that $\Delta n(x)$ has the same periodicity and shape as $V^{\rm im}(x)$, oscillating between the values $n_1$ and $-n_1$ for $b=a/2$ so that the average electron density is conserved. By specifying Eq. (\ref{lagmult}) at two different points $0<x_1<b$ and $b<x_2<a$, and then subtracting the two resulting equations, we find the ratio
\begin{align}
\frac{n_1}{n_0}=\frac{1}{2}\frac{V_0/\EF^0}{1+\mathcal{A}[(V_{\rm C}/\EF^0)/(k_{\rm BZ}/\kF^0)^2]},
\end{align}
where $\mathcal{A}=\sum_{G\neq 0}(\pi^2/4a|G|^2)K_G(\ee^{\ii G x_1 }-\ee^{\ii G x_2})$, $K_G=\Delta n_G /n_1=(2\ii/a G)(\ee^{-\ii G b}-1)$, and we use the coefficients $V_{\rm C}=e^2/a$ and $k_{\rm BZ}=\pi/a$ defined in the main text. We have verified that $\mathcal{A}$ evolves in the $(0,1)$ interval as the values of $x_1<x_2$ are varied.  For instance, if $x_1=a/4$ and $x_2=3a/4$, we obtain $\mathcal{A}=\sum_{n=0}^\infty(-1)^n/(2n+1)^2\sim 0.91$. We are interested in finding the ratio of the kinetic energy to the image potential energy, which in this approximation becomes
\begin{align}
\left|\frac{E^{\rm kin}}{E^{\rm im}}\right|&=\frac{\pi\hbar^2}{2m^*}\frac{\int dx\, n^2(x)}{\left|\int dx\, V^{\rm im}(x)n(x)\right|} \nonumber\\
&=\frac{\EF^0}{2 V_0}\frac{1+n_1^2/n_0^2}{|1+n_1/n_0|}, \nonumber
\end{align}
and finally, for small perturbations ($|n_1|\ll n_0$), it reduces to
\begin{align}
 \left|\frac{E^{\rm kin}}{E^{\rm im}}\right| \approx \left|\frac{\EF^0}{2V_0} - \frac{1/4}{1+\mathcal{A}[(V_{\rm C}/\EF^0)/(k_{\rm BZ}/\kF^0)^2]}\right|. 
 \end{align}
The second fraction in the right-hand side of this equation can be neglected under the conditions investigated in the main text (i.e., for $V_{\rm C}/\EF^0 /( k_{\rm BZ}/\kF^0)^2\gg 1$), so the influence of the image potential on the material is simply quantified through the parameter $V_0/\EF^0$.

\section{Optical response in the RPA}\label{meth2}

We start by writing the charge distribution $\rho^{\rm ind}$ induced in the semiconductor layer in the presence of an external electric potential $\phi^{\rm ext}$ as $\rho^{\rm ind}=\chi\phi^{\rm ext}$, where $\chi$ is the electric susceptibility and an overall $\ee^{-\ii\omega t}$ time dependence is understood. We now adopt the RPA \cite{HL1970} to write $\chi=\chi^0(1-\upsilon\chi^0)^{-1}$ in terms of the noninteracting susceptibility $\chi^0(\rb,\rb',\omega)=\delta(z)\delta(z')\tilde{\chi}^0(\Rb,\Rb',\omega)$, where
\begin{align}
&\tilde{\chi}^0(\Rb,\Rb',\omega)=\frac{2e^2}{\hbar}\sum_{\kparb\kparb',nn'}\,\frac{f_{\kparb' n'}-f_{\kparb n}}{\omega -\varepsilon_{\kparb n}+\varepsilon_{\kparb' n'}+\ii0^+} \nonumber\\
\times\,  &\psi_{\kparb n}(\Rb)\psi^*_{\kparb' n'}(\Rb)\psi_{\kparb' n'}(\Rb')\psi_{\kparb n}^*(\Rb'), \label{chi0}
\end{align}
and obviously, the full susceptibility also bears an out-of-plane dependence as $\chi(\rb,\rb',\omega)=\delta(z)\delta(z')\tilde{\chi}(\Rb,\Rb',\omega)$. Here, $\upsilon(\rb,\rb')$ is the Coulomb potential produced at $\rb$ by a unit charge placed at $\rb '$, including the effect of screening by the surrounding environment. For simplicity, we assume the material in the ribbon array to be perfectly conducting at zero frequency, but invisible at the infrared optical frequencies under consideration (e.g., by embedding the entire structure in an index-matching medium), so that $\upsilon$ can be well approximated by the bare Coulomb potential $\upsilon(\rb,\rb')\approx 1/|\rb-\rb '|$ in the calculation of $\chi$.  Now, given the ribbon array periodicity, we have $\tilde{\chi}(\Rb,\Rb ',\omega)=\tilde{\chi}(\Rb+\xx \ell a,\Rb '+\xx \ell a,\omega)$ for any integer $\ell$, so we can write $\tilde{\chi}(\Rb,\Rb ',\omega)=(2\pi)^{-2}\sum_{GG'}\ee^{\ii(Gx-G'x')}\int_{-\pi/a}^{\pi/a} dq_x\int_{-\infty}^\infty dq_y\,\tilde{\chi}_{GG'}(\qparb,\omega)$ $\times\ee^{\ii\qparb\cdot(\Rb-\Rb')}$, where the $q_x$ integral is limited to the 1D 1BZ of the ribbon lattice and the components $\tilde{\chi}_{GG'}(\qparb,\omega)$ are all we need to calculate the optical response to an external perturbation bearing an in-plane spatial dependence $\propto\ee^{\ii\qparb\cdot\Rb}$. Finally, inserting the one-electron wave functions discussed above into Eq.\ (\ref{chi0}), we readily find Eq.\ (\ref{chi0main}) in the main text.

\section{Thermal and electrical conductivities}\label{meth4}

An external static in-plane electric field $\Eb$ produces a 2D current density $\jb^{\rm e}=\sigma^{\rm DC} \Eb$, where $\sigma^{\rm DC}$ is the local DC ($\omega=0$) electrical conductivity tensor. Likewise, an in-plane temperature gradient induces a 2D thermal current density $\jb^{\rm th}=- \sigma^{\rm th} \nabla_{\Rb}T$, where $\sigma^{\rm th}$  is the electronic thermal conductivity tensor. In the relaxation-time approximation, the effect of inelastic electron scattering enters via a phenomenological energy-independent damping time $\tau$, contributed by different scattering channels such as acoustic phonons \cite{HS08,KTJ12}, and we find the $2\times2$ tensors \cite{AM1976}
\begin{align}
\sigma^{\rm DC}=&\frac{e^2 \tau}{2\pi^2} \sum_n \int_{-\pi/a}^{\pi/a} d\kx \int_{-\infty}^\infty d\ky\, \delta(\EF-\hbar \varepsilon_{\kparb n})\, \vb_{\kparb n}\otimes \vb_{\kparb n},\nonumber\\
\sigma^{\rm th}=&\frac{\pi^2 k_{\rm B}^2T}{3e^2}\left[\sigma^{\rm DC} - \frac{\pi^2k_{\rm B}^2T^2}{3} \,\sigma'\cdot\big(\sigma^{\rm DC}\big)^{-1}\cdot\sigma'\right],\nonumber
\end{align}
where $\vb_{\kparb n}=\nabla_{\kparb}\varepsilon_{\kparb n}$ is the electron velocity and $\sigma' = (e^2\tau/2\pi^2)\sum_n \int_{-\pi/a}^{\pi/a} d\kx \int_{-\infty}^\infty d\ky \, \delta(\EF-\hbar \varepsilon_{\kparb n}) \,  (m_{\kparb n}^*)^{-1}$ is expressed in terms of the effective mass tensor $ m_{\kparb n}^*=\hbar [(\nabla_{\kparb n}\otimes \nabla_{\kparb n})\varepsilon_{\kparb n}]^{-1}$ \cite{AM1976}. Equations~(\ref{sigdc}) and (\ref{sigth}) in the main text follow directly from these expressions by using the electron dispersion in Eq.~(\ref{eigenval}).

\begin{figure}
\centering \includegraphics[width=0.45\textwidth]{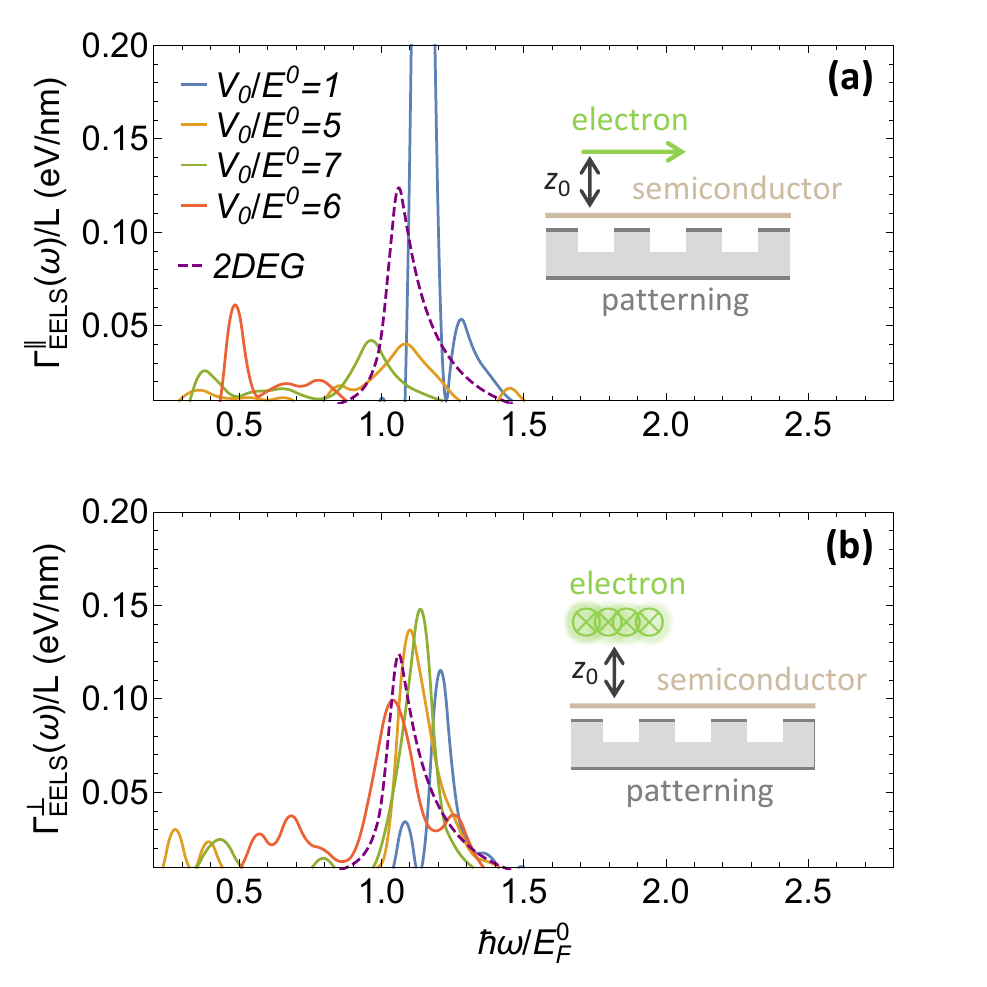}
\caption{\textbf{Probing the optical response of a Q-phase material through EELS.} (a) EELS probability experienced by a 50\,eV electron moving parallel to the semiconductor under consideration at a distance $z_0=10$\,nm (see inset). We show results for different values of $V_0/\EFz$ (solid curves, obtained by inserting the loss function of Fig.~\ref{Fig2} into Eq.\ (\ref{EELSpara})), compared with the probability calculated for a 2DEG described through the local Drude conductivity $\sigma(\omega)=(4\pi\ii n_0/m^*)\big/(\omega +\ii /\tau)$ (dashed curve). A phenomenological damping $\hbar/\tau=10$\,meV is assumed in all cases. (b) Same as (a), but with the electron traveling parallel to the ribbons, and the probability averaged over lateral position $x$ according to Eq.\ (\ref{EELSperp}).}
\label{FigS1}
\end{figure}

\section{Probing the optical response through EELS}

The optical response of the Q-phase materials under consideration can be probed through electron energy-loss spectroscopy (EELS). We follow the general methods discussed elsewhere \cite{paper149} to calculate the loss probability for an electron moving with constant velocity $\vb$ parallel to the semiconductor and oriented along in-plane directions either parallel or perpendicular to the ribbons. The electron is taken to be moving in vacuum on the side of the semiconductor that is not occupied by the ribbon structure used to produce the image potential (see insets in Fig.~\ref{FigS1}).

Adopting the electrostatic limit, the EELS probability $\Gamma_{\rm EELS}(\omega)$ is directly obtained from the screened Coulomb interaction $W(\rb,\rb',\omega)$, which describes the potential produced at $\rb$ by a unit point charge placed at $\rb'$ and oscillating with frequency $\omega$. More precisely \cite{paper149},
\begin{align}
\Gamma_{\rm EELS}(\omega)=&\frac{e^2}{\pi \hbar} \int dt \int dt'\, \ee^{\ii \omega (t'-t)} \label{EELSW}\\
&\times{\rm Im}\left\{-W^{\rm ind}(\rb_0+\vb t,\rb_0+\vb t',\omega)\right\}, \nonumber
\end{align}
where $\rb_0$ describes the electron position at time $t=0$, and $z_0$ is the electron-surface separation. Now, from the analysis presented in the main text, defining $\Gb=G\xx$ with $G=2\pi m/a$ and $m$ running over integer numbers, we consider points $z,z'>0$ above the semiconductor surface and express the screened interaction in terms of the Fourier components of the Fresnel reflection coefficient for p polarization:
\begin{align}
&W^{\rm ind}(\rb,\rb',\omega)=-\frac{1}{2\pi}\int_{-\pi/a}^{\pi/a}\!\!dq_x\int_{-\infty}^\infty\!\!dq_y\sum_{GG'}\frac{r^{\rm p}_{GG'}(\qparb,\omega)}{|\qparb+\Gb'|} \nonumber\\
&\times\ee^{\ii (\qparb+\Gb)\cdot\Rb}\ee^{-\ii (\qparb+\Gb')\cdot\Rb'}\ee^{-|\qparb+\Gb|z}\ee^{-|\qparb+\Gb'|z'}.  \label{W}
\end{align}
Incidentally, we only write the induced part of the interaction because the direct Coulomb term does not contribute to Eq.\ (\ref{EELSW}).

For an electron beam oriented along $y$ (parallel to the ribbons), we take $\rb_0=(x_0,0,z_0)$ and average over lateral beam positions $x_0$ across one period of the array, so that by inserting Eq.\ (\ref{W}) into Eq.\ (\ref{EELSW}) and using the integral $(1/a)\int_0^a dx_0\,\ee^{\ii(G-G')x_0}=\delta_{GG'}$, we find
\begin{align}
\Gamma^\perp_{\rm EELS}(\omega)=\frac{2e^2L}{\pi\hbar v^2} &\int_0^{\pi/a}\!\!dq_x \sum_G
\frac{\ee^{-2Q_1z_0}}{Q_1} \label{EELSperp}\\
&\times{\rm Im}\left\{r^{\rm p}_{GG}(q_x+G,\omega/v,\omega)\right\}, \nonumber
\end{align}
where $Q_1=\sqrt{(q_x+G)^2+\omega^2/v^2}$, $L$ is the length of the electron trajectory, and the $\perp$ superscript refers to the fact that the beam is moving perpendicularly with respect to the direction of the array periodicity.

Likewise, for an electron traveling along the transverse ribbon direction $x$ (i.e., parallel to the array periodicity), we can take $\rb_0=(0,0,z_0)$ and insert Eq.\ (\ref{W}) into Eq.\ (\ref{EELSW}) again to readily obtain
\begin{align}
\Gamma^\parallel_{\rm EELS}(\omega)=\frac{2e^2L}{\pi\hbar v^2} &\int_0^\infty dq_y
\frac{\ee^{-2Q_2z_0}}{Q_2} \label{EELSpara}\\
&\times{\rm Im}\left\{r^{\rm p}_{\tilde{G}\tilde{G}}(\tilde{q}_x,q_y,\omega)\right\}, \nonumber
\end{align}
where $Q_2=\sqrt{\omega^2/v^2+q_y^2}$, $\tilde{q}_x=\omega/v-\tilde{G}$, and $\tilde{G}$ is the only lattice vector satisfying the condition $|\tilde{q}_x|<\pi/a$. Obviously, $\tilde{G}$ may depend on $\omega/v$, although we expect to have $\omega/v\ll\pi/a$, and therefore $G=0$. In such scenario, Eq.\ (\ref{EELSpara}) coincides with the electrostatic limit of the probability obtained for an electron moving parallel to the surface of a photonic crystal \cite{paper045}.

In Fig.~\ref{FigS1}, we present EELS spectra calculated by using Eqs.~(\ref{EELSperp}) and (\ref{EELSpara}) for different strengths of the image interaction $V_0$ down to the 2DEG limit $V_0=0$ (dashed curves). Strong deviations are observed in the number and positions of the spectral features inherited from the dispersion relations presented in Fig.~\ref{Fig2}. The in-plane anisotropy of the designed Q-phase material gives rise to substantially different profiles for the two electron beam orientations under consideration.



\section*{Acknowledgments} 
This work has been supported in part by the European Research Council (Advanced Grant 789104-eNANO), the Spanish MICINN (PID2020-112625GB-I00 and Severo Ochoa CEX2019-000910-S), the Catalan CERCA Program, and Fundaci\'{o}s Cellex and Mir-Puig. 


\end{document}